\documentclass[aps,prb,twocolumn]{revtex4-1}

\usepackage{graphicx}
\usepackage{dcolumn}
\usepackage{bm}
\usepackage[version=3]{mhchem}

\begin{document}

\title{Understanding the electronic and ionic conduction and lithium over-stoichiometry in \ce{LiMn2O4} spinel}
\author{Khang Hoang}
\affiliation{Center for Computationally Assisted Science and Technology, North Dakota State University, Fargo, ND 58108, USA.}


\begin{abstract}

We report a first-principles study of defect thermodynamics and transport in spinel-type lithium manganese oxide \ce{LiMn2O4}, an important lithium-ion battery electrode material, using density-functional theory and the Heyd-Scuseria-Ernzerhof screened hybrid functional. We find that intrinsic point defects in \ce{LiMn2O4} have low formation energies and hence can occur with high concentrations. The electronic conduction proceeds via hopping of small polarons and the ionic conduction occurs via lithium vacancy and/or interstitialcy migration mechanisms. The total conductivity is dominated by the electronic contribution. \ce{LiMn2O4} is found to be prone to lithium over-stoichiometry, {\it i.e.}, lithium excess at the manganese sites, and Mn$^{3+}$/Mn$^{4+}$ disorder. Other defects such as manganese antisites and vacancies and lithium interstitials may also occur in \ce{LiMn2O4} samples. In light of our results, we discuss possible implications of the defects on the electrochemical properties and provide explanations for the experimental observations and guidelines for defect-controlled synthesis and defect characterization.

\end{abstract}


\maketitle


\section{Introduction}

Spinel-type \ce{LiMn2O4}, a mixed-valent compound containing both Mn$^{3+}$ and Mn$^{4+}$ ions, has been considered as an alternative to layered \ce{LiCoO2} for lithium-ion battery electrodes as manganese is inexpensive and environmentally benign compared to cobalt in the layered oxide.\cite{Thackeray1999,Park2011} The material crystallizes in the cubic crystal structure of space group $Fd\overline{3}m$ at room temperature but transforms into an orthorhombic or tetragonal structure at lower temperatures. In the cubic phase, the Li$^{+}$ ions stay at the tetrahedral 8a sites of the cubic close-packed oxygen array, whereas the Mn$^{3+}$ and Mn$^{4+}$ ions randomly occupy the octahedral 16d sites. In the orthorhombic or tetragonal phase, the Mn$^{3+}$/Mn$^{4+}$ arrangement is believed to be charge-ordered.\cite{Carvajal1998,Piszora2004,Oikawa1998,Akimoto2004,Wills1999,Yamada1995,Yonemura2004} However, truly stoichiometric \ce{LiMn2O4} is hard to obtain in experiments and intrinsic electronic and ionic defects appear to occur at multiple lattice sites.\cite{Tarascon1994,Bjork2001,Yonemura2004,Strobel2004,Martinez2014,Paulsen1999,Lee2002} In fact, \ce{LiMn2O4} samples are often made lithium over-stoichiometric ({\it i.e.}, Li-excess), intentionally or unintentionally.\cite{Paulsen1999,Lee2002,Takada1999,Martinez2014,Berg1999,Sugiyama2007,Kamazawa2011,Gummow1994,Xia1996,Massarotti1999} The total bulk conductivity of \ce{LiMn2O4} has also been reported and thought to be predominantly from hopping of polarons.\cite{Massarotti1999,Iguchi2002,Fang2008} A deeper understanding of these properties and observations clearly requires a detailed understanding of the defect thermodynamics and transport. Such an understanding is currently lacking.

Computational studies of \ce{LiMn2O4} have focused mainly on the bulk properties and lithium and small polaron migration,\cite{Grechnev2002,Wang2007,Ouyang2009PLA,Nakayama2010,Xu2010,Ammundsen1997} except for some studies of defect energetics by Ammundsen {\it et al.}\cite{Ammundsen1997} using interatomic potential simulations and Koyama {\it et al.}\cite{Koyama2003} using density-functional theory (DFT) within the local density approximation (LDA).\cite{LDA} First-principles calculations based on DFT have been proven to be a powerful tool for addressing electronic and atomistic processes in solids. A comprehensive and systematic DFT study of intrinsic point defects in a battery electrode material, for example, can provide a detailed picture of the defect formation and migration and invaluable insights into the electrochemical performance.\cite{Hoang2011,Hoang2014} For \ce{LiMn2O4}, such a study is quite challenging, partly because standard DFT calculations using LDA or the generalized-gradient approximation (GGA)\cite{GGA} fail to produce the correct physics of even the host compound. LDA/GGA calculations carried out by Mishra and Ceder,\cite{Mishra1999} {\it e.g.}, showed that \ce{LiMn2O4} is a metal with a Mn oxidation state of +3.5, which is in contrast to what is known about \ce{LiMn2O4} as a material with a finite band gap and mixed Mn$^{3+}$ and Mn$^{4+}$ ions. The GGA$+U$ method,\cite{dudarev1998,liechtenstein1995} an extension of GGA, can give a reasonable electronic structure. However, since one often has to assume that the transition metal has the same Hubbard $U$ value in different chemical environments, the transferability of GGA$+U$ results across the compounds is low, making defect calculations become inaccurate.

In this article, we present for the first time a comprehensive study of electronic and ionic defects in \ce{LiMn2O4} using a hybrid Hartree-Fock/DFT method. In particular, we used the Heyd-Scuseria-Ernzerhof (HSE06)\cite{heyd:8207,paier:154709} screened hybrid functional where all orbitals are treated on the same footing. The atomic and electronic structure and phase stability of the host compound and the structure and energetics of all possible intrinsic point defects were investigated; the migration of selected defects was also explored. We find that defects in \ce{LiMn2O4} have low calculated formation energies and hence can occur with high concentrations, and lithium antisites are the dominant ionic defect. On the basis of our results, we discuss the Mn$^{3+}$/Mn$^{4+}$ disorder, electronic and ionic conduction, delithiation and lithiation mechanisms, lithium over-stoichiometry, and possible implications on the electrochemical properties. Ultimately, our work provides explanations for the experimental observations, guidelines for defect characterization and defect-controlled synthesis, and insights for rational design of \ce{LiMn2O4}-based electrode materials with improved electrochemical performance.

\section{Methodology}

\subsection{Computational details} 

Our calculations were based on DFT, using the HSE06 hybrid functional,\cite{heyd:8207,paier:154709} the projector augmented wave method,\cite{PAW1,PAW2} and a plane-wave basis set, as implemented in the Vienna {\it Ab Initio} Simulation Package (VASP).\cite{VASP1,VASP2,VASP3} Point defects were treated within the supercell approach, in which a defect is included in a finite volume of the host material and this structure is periodically repeated. For bulk and defect calculations, we mainly used supercells of \ce{LiMn2O4} containing 56 atoms/cell; integrations over the Brillouin zone were carried out using a 2$\times$2$\times$2 Monkhorst-Pack $\mathbf{k}$-point mesh.\cite{monkhorst-pack} A denser, $\Gamma$-centered 4$\times$4$\times$4 $\mathbf{k}$-point mesh was used in calculations to produce the electronic density of states. The plane-wave basis-set cutoff was set to 500 eV. Convergence with respect to self-consistent iterations was assumed when the total energy difference between cycles was less than 10$^{-4}$ eV and the residual forces were less than 0.01 eV/{\AA}. In the defect calculations, which were performed with spin polarization and the ferromagnetic spin configuration, the lattice parameters were fixed to the calculated bulk values but all the internal coordinates were fully relaxed.

\subsection{Defect formation energies}

The key quantities that determine the properties of a defect are the migration barrier and formation energy. In our calculations, the former is calculated by using climbing-image nudged elastic-band (NEB) method;\cite{ci-neb} the latter is computed using the total energies from DFT calculations. Following the approach described in Ref.\cite{Hoang2014} and references therein, the formation energy of a defect X in charge state $q$ is defined as
\begin{equation}\label{eq;eform}
E^f({\mathrm{X}}^q)=E_{\mathrm{tot}}({\mathrm{X}}^q)-E_{\mathrm{tot}}({\mathrm{bulk}})-\sum_{i}{n_i\mu_i}+q(E_{\mathrm{v}}+\mu_{e})+ \Delta^q,
\end{equation}
where $E_{\mathrm{tot}}(\mathrm{X}^{q})$ and $E_{\mathrm{tot}}(\mathrm{bulk})$ are, respectively, the total energies of a supercell containing the defect X and of a supercell of the perfect bulk material; $\mu_{i}$ is the atomic chemical potential of species $i$ (and is referenced to bulk metals or O$_{2}$ molecules at 0 K), and $n_{i}$ indicates the number of atoms of species $i$ that have been added ($n_{i}$$>$0) or removed ($n_{i}$$<$0) to form the defect. $\mu_{e}$ is the electronic chemical potential, referenced to the valence-band maximum in the bulk ($E_{\mathrm{v}}$). $\Delta^q$ is the correction term to align the electrostatic potentials of the bulk and defect supercells and to account for finite-cell-size effects on the total energies of charged defects.\cite{walle:3851} To correct for the finite-size effects, we adopted the approach of Freysoldt {\it et al.},\cite{Freysoldt,Freysoldt11} in which $\Delta^q$ was determined using a calculated static dielectric constant of 11.02 for \ce{LiMn2O4}. The dielectric constant was computed following the procedure described in Ref.\cite{Hoang2014} according to which the electronic contribution (4.78) was obtained in HSE06 calculations whereas the ionic contribution (6.24) was obtained in GGA+$U$ with $U$=4.84 eV for Mn, taken as an average value of Mn$^{3+}$ (4.64 eV) and Mn$^{4+}$ (5.04 eV).\cite{Zhou:2004p104}

In eqn~(\ref{eq;eform}), the atomic chemical potentials $\mu_{i}$ can be employed to describe experimental conditions and are subject to various thermodynamic constraints.\cite{Hoang2014,walle:3851} The stability of \ce{LiMn2O4}, for example, requires
\begin{equation}\label{eq;stability} 
\mu_{\rm Li}+2\mu_{\rm Mn}+4\mu_{\rm O}=\Delta H^{f}({\rm LiMn}_{2}{\rm O}_{4}), 
\end{equation} 
where $\Delta H^{f}$ is the formation enthalpy. There are other constraints imposed by competing Li$-$Mn$-$O phases. By taking into account all these thermodynamic constraints, one can determine the range of Li, Mn, and O chemical potential values in which the host compound \ce{LiMn2O4} is thermodynamically stable. The oxygen chemical potential, $\mu_{\rm O}$, can also be related to the temperatures and pressures through standard thermodynamic expressions for \ce{O2} gas.\cite{Hoang2011} Finally, the electronic chemical potential $\mu_{e}$, {\it i.e.}, the Fermi level, is not a free parameter but subject to the charge neutrality condition that involves all possible intrinsic defects and any impurities in the material.\cite{Hoang2014,walle:3851}

The concentration of a defect at temperature $T$ is related to its formation energy through the expression\cite{walle:3851} 
\begin{equation}\label{eq;concen} 
c=N_{\mathrm{sites}}N_{\mathrm{config}}\mathrm{exp}\left(\frac{-E^{f}}{k_{B}T}\right), 
\end{equation} 
where $N_{\mathrm{sites}}$ is the number of high-symmetry sites in the lattice per unit volume on which the defect can be incorporated, $N_{\mathrm{config}}$ is the number of equivalent configurations (per site), and $k_{B}$ is the Boltzmann constant. Strictly speaking, this expression is only valid in thermodynamic equilibrium. Materials synthesis, on the other hand, may not be an equilibrium process. However, even in that case the use of the equilibrium expression can still be justified if the synthesis conditions are close enough to equilibrium. Besides, as discussed in Ref.\cite{walle:3851}, the use of eqn~(\ref{eq;concen}) does not require that all aspects of the process have to be in equilibrium. What is important is that the relevant defects are mobile enough to allow for equilibration at the temperatures of interest. It emerges from eqn~(\ref{eq;concen}) that defects with low formation energies will easily form and occur in high concentrations.

\section{Results}
\subsection{Bulk properties}

\begin{figure}
\centering
\includegraphics[width=3.2in]{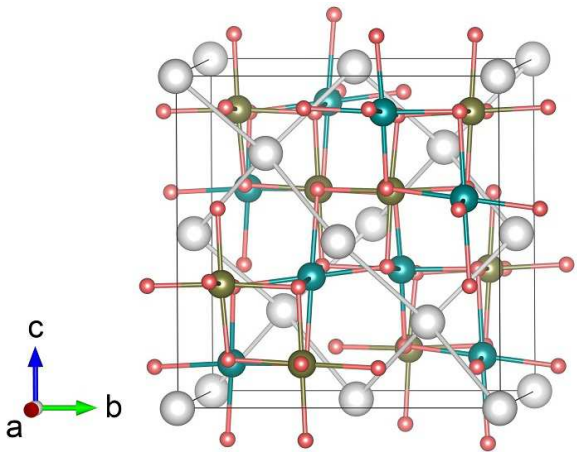}
\caption{Supercell model for spinel-type \ce{LiMn2O4} after structural optimization. Large gray spheres are Li, medium blue (yellow) spheres are Mn$^{3+}$ (Mn$^{4+}$), and small red spheres are O.}
\label{fig;struct}
\vspace{0.3cm}
\end{figure}

We began with a cubic supercell of \ce{LiMn2O4} (space group $Fd\overline{3}m$, experimental lattice parameter $a = 8.24$ {\AA}),\cite{Mukai2011} consisting of 8 Li atoms at the 8a sites, 16 Mn atoms at the 16d sites, and 32 O atoms at the 32e sites; the interstitial 16c sites are left empty. After structural optimization, this cubic cell transforms into a tetragonally distorted cell with $a=c=8.34$ {\AA} and $b=8.11$ {\AA; see Fig.~\ref{fig;struct}. The ordering of Mn$^{3+}$/Mn$^{4+}$ is visible in the relaxed structure. If viewed along the [101] direction, the atomic arrangement follows an A$-$B$-$C$-$... pattern with layer A consisting of Mn$^{3+}$ chains, B of Li$^{+}$ chains and Mn$^{3+}$/Mn$^{4+}$-alternating chains, and C of Mn$^{4+}$ chains. The Mn ions are stable in high-spin states with calculated magnetic moments of 3.78 $\mu_{\rm B}$ (Mn$^{3+}$) and 3.04 $\mu_{\rm B}$ (Mn$^{4+}$). The MnO$_6$ unit has either six Mn$^{4+}$$-$O bonds with similar bond lengths (1.85$-$1.93 {\AA}) or four short Mn$^{3+}$$-$O bonds (1.92$-$1.98 {\AA}) and two long Mn$^{3+}-$O bonds (2.20 {\AA}). The local distortions as seen in Fig.~\ref{fig;struct} are thus due to Jahn-Teller effects associated with the Mn$^{3+}$ ions. 

Among several different Mn$^{3+}$/Mn$^{4+}$ arrangements we investigated, the described model is found to have the lowest energy. It is lower in energy than the second-lowest energy Mn$^{3+}$/Mn$^{4+}$ arrangement by 0.06 eV per formula unit (f.u.), and the \ce{LiMn2O4} supercell where every Mn ion has an oxidation state of +3.5 by 0.73 eV per f.u. Our results are thus consistent with experimental reports showing a transformation into a tetragonal or orthorhombic phase at low temperatures associated with charge ordering.\cite{Carvajal1998,Piszora2004,Oikawa1998,Akimoto2004,Wills1999,Yamada1995,Yonemura2004} We use this structural model for further studies of the bulk properties and for defect calculations (see below). Since the above mentioned global and local distortions are relatively small, the atomic positions in this model will be nominally referred to using the Wyckoff positions of the cubic structure.

\begin{figure}
\vspace{0.2cm}
\centering
\includegraphics[width=3.2in]{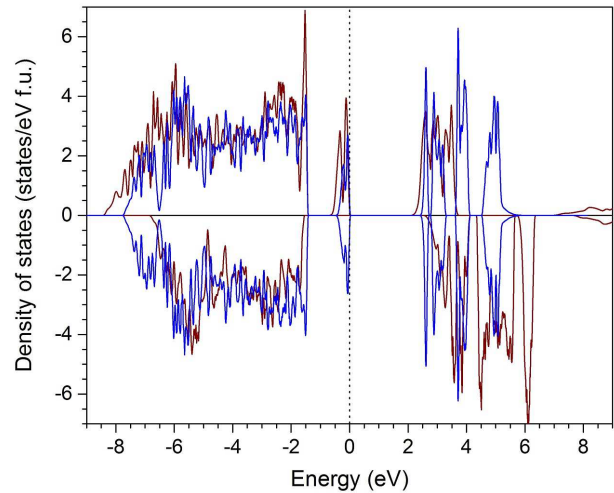}
\caption{ Electronic density of states of \ce{LiMn2O4} in antiferromagnetic (blue curves) and ferromagnetic (red curves) spin configurations. The zero of energy is set to the highest occupied state.}
\label{fig;dos}
\vspace{0.3cm}
\end{figure}

Figure~\ref{fig;dos} shows the total electronic density of states of \ce{LiMn2O4}. An analysis of the wavefunctions shows that the valence-band maximum (VBM) predominantly consists of the 3d states from the Mn$^{3+}$ sites, whereas the conduction-band minimum (CBM) are predominantly the 3d states from the Mn$^{4+}$ sites. The Li 2s state is high up in the conduction band, indicating that Li donates its electron to the lattice and becomes Li$^{+}$. \ce{LiMn2O4} thus can be regarded nominally as an ordered arrangement of Li$^{+}$, Mn$^{3+}$, Mn$^{4+}$, and O$^{2-}$ units. The calculated band gaps are 2.12 and 2.42 eV for the ferromagnetic (FM) and antiferromagnetic (AFM) configurations, respectively. In addition to the spin configurations, we find that the calculated band gap of \ce{LiMn2O4} also depends on the Mn$^{3+}$/Mn$^{4+}$ arrangement. For example, our calculations using a smaller, 14-atom cell of \ce{LiMn2O4}, which also relaxes to a tetragonal structure but with a different Mn$^{3+}$/Mn$^{4+}$ arrangement, give band gaps of 1.77 and 1.92 eV for the FM and AFM configurations. 

Experimentally, Raja {\it et al.}\cite{Raja2007} reported an optical band gap of 1.43 eV from ultraviolet-visible spectroscopy for nanocrystalline \ce{LiMn2O4} powders with a nominal composition of Li$_{0.88}$Mn$_2$O$_4$. Kushida and Kuriyama,\cite{Kushida2000} on the other hand, observed two optical absorption peaks associated with d$-$d transitions at about 1.63 and 2.00 eV in \ce{LiMn2O4} thin films on silica glass. The discrepancies in the experimental values suggest that the band gap value is sensitive to the quality of the \ce{LiMn2O4} samples, which in turn depends on the synthesis conditions. This obviously complicates the comparison between the calculated and measured bulk properties.

\subsection{Chemical-potential diagram}

\begin{figure}
\centering
\includegraphics[width=3.2in]{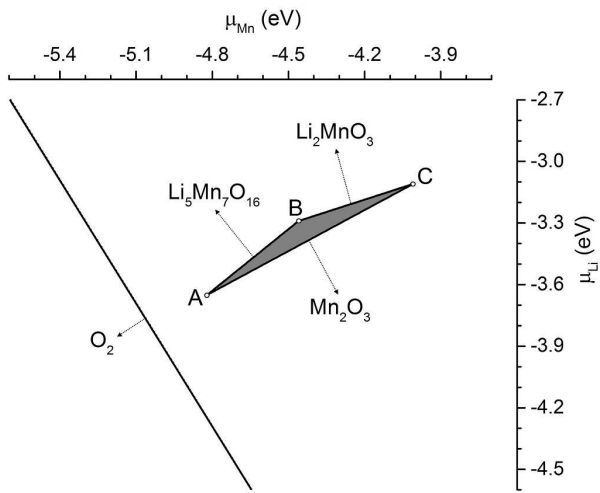}
\caption{Chemical-potential diagram at 0 K for \ce{LiMn2O4}. Only the \ce{O2} gas phase and the Li$-$Mn$-$O phases that define the stability region, here shown as a shaded triangle, are included.}
\label{fig;chempot}
\vspace{0.3cm}
\end{figure}

Figure~\ref{fig;chempot} shows the atomic chemical-potential diagram for \ce{LiMn2O4}, constructed by exploring all possible Li$-$Mn$-$O phases available in the Materials Project database.\cite{Jain2013} The stability region of \ce{LiMn2O4} in the ($\mu_{\rm Li}$, $\mu_{\rm Mn}$) plane is defined by \ce{Mn2O3}, \ce{Li2MnO3}, and \ce{Li5Mn7O16}. The calculated formation enthalpies at 0 K of tetragonal \ce{LiMn2O4}, orthorhombic \ce{Mn2O3}, monoclinic \ce{Li2MnO3}, and orthorhombic \ce{Li5Mn7O16} are, respectively, $-$13.89 eV, $-$10.09 eV, $-$12.30 eV, and $-$54.39 eV/f.u. For comparison, the experimental formation enthalpy of \ce{LiMn2O4} at 298 K is $-$14.31 eV/f.u.\cite{Wang20051182} Points A, B, and C represent three-phase equilibria associated with \ce{LiMn2O4}. Point A, for example, is an equilibrium between \ce{LiMn2O4}, \ce{Mn2O3}, and \ce{Li5Mn7O16}. The presence of these equilibria is consistent with the fact that \ce{LiMn2O4} samples often contain \ce{Mn2O3} and/or \ce{Li2MnO3} as impurity phases.\cite{Berg1999,Gao1996,Luo2007} Strobel {\it et al.}\cite{Strobel2004} also reported that annealing \ce{LiMn2O4} under oxygen pressures in the range 0.2$-$5 atm at 450$^\circ$C resulted in Mn atoms being expelled in form of \ce{Mn2O3}. \ce{Li5Mn7O16}, which can be rewritten as Li$_{1+\alpha}$Mn$_{2-\alpha}$O$_{4}$ ($\alpha=0.25$), is also closely related to \ce{LiMn2O4}. In fact, in the Li$-$Mn$-$O phase diagram, it is located on the tie-line between \ce{LiMn2O4} ($\alpha=0$) where the average Mn oxidation state is +3.5 and \ce{Li4Mn5O12} ($\alpha=0.33$) where all Mn ions have the oxidation state of +4. 

\subsection{Defect structure and energetics}

\begin{figure}
\centering
\includegraphics[width=3.2in]{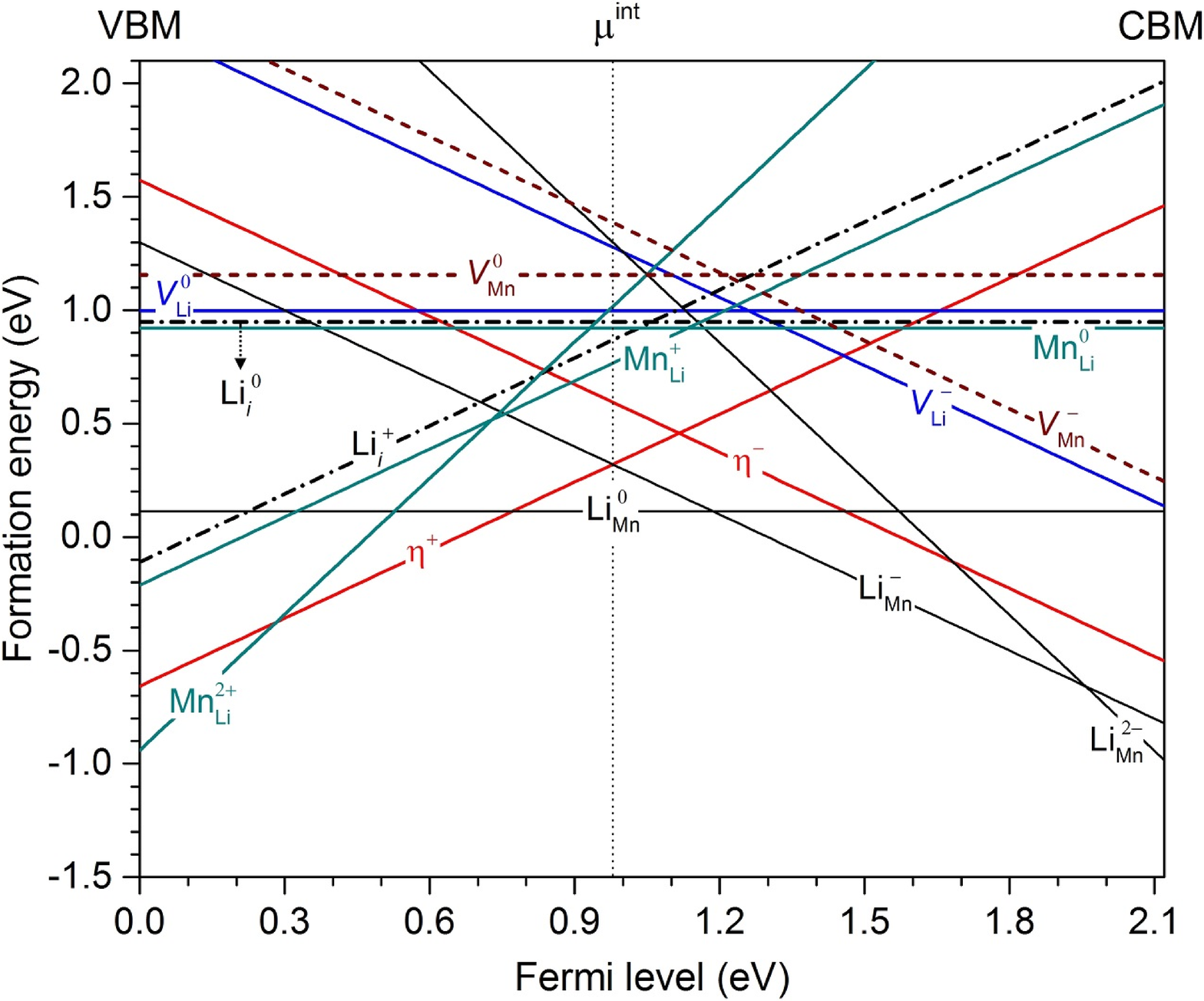}
\caption{Calculated formation energies of intrinsic point defects in \ce{LiMn2O4}, plotted as a function of the Fermi level. The energies are obtained at point B in Fig.~\ref{fig;chempot}. In the absence of extrinsic charged impurities, the Fermi level is at $\mu_{e}=\mu_{e}^{\rm{int}}$, where charge neutrality is maintained.}
\label{fig;formenergy}
\end{figure}

Figure~\ref{fig;formenergy} shows the calculated formation energies of low-energy defects in \ce{LiMn2O4}, obtained at point B in the chemical-potential diagram. These defects include hole ($\eta^{+}$) and electron ($\eta^{-}$) polarons, lithium vacancies ($V_{\rm Li}$), interstitials (Li$_{i}$), and antisites (Li$_{\rm Mn}$), and manganese vacancies ($V_{\rm Mn}$) and antisites (Mn$_{\rm Li}$). The formation energies are plotted as a function of the Fermi level $\mu_{e}$, with $\mu_{e}$ varies from the VBM to CBM. As mentioned earlier, the actual position of the Fermi level of the system is determined by the charge neutrality condition. In the absence of electrically active impurities that can shift the Fermi-level position or when such impurities occur in much lower concentrations than charged intrinsic defects, the Fermi level is at $\mu_{e}^{\rm int}$, determined only by the intrinsic defects.\cite{Hoang2011,Hoang2014} With the chosen set of the atomic chemical potentials, $\mu_{e}^{\rm int}$ is at 0.98 eV, exclusively defined by small hole polarons ($\eta^{+}$) and negatively charged lithium antisites (Li$_{\rm Mn}^{-}$). We find that intrinsic point defects in \ce{LiMn2O4} have very low formation energies and hence can occur with high concentrations. Polarons and charged lithium and manganese antisites have positive calculated formation energies only near midgap. Positively charged lithium interstitials (Li$_{i}^{+}$) also have a negative formation energy near the VBM. Before discussing the implications of these results, let us describe the defects in more detail.

{\bf Small polarons.} A hole (or electron) polaron is a quasiparticle formed by the hole (electron) and its self-induced local lattice distortion. The creation of $\eta^{+}$ involves removing an electron from the VBM which is predominantly Mn$^{3+}$ 3d states. This results in a Mn$^{4+}$ ion at one of the Mn$^{3+}$ sites, {\it i.e.}, a localized hole. The local lattice geometry near the newly formed Mn$^{4+}$ ion is slightly distorted with the six neighboring O atoms moving toward the Mn$^{4+}$. The average Mn$-$O bond length at the Mn$^{4+}$ site is 1.92 {\AA} and the Jahn-Teller distortion vanishes at this site. The formation of $\eta^{-}$, on the other hand, corresponds to adding an electron to the CBM, which is predominantly Mn$^{4+}$ 3d states, resulting in a Mn$^{3+}$ ion at one of the Mn$^{4+}$ sites, {\it i.e.}, a localized electron. The local geometry near the newly formed Mn$^{3+}$ ion is also distorted, but with the neighboring O atoms slightly moving away from Mn$^{3+}$. At this Mn$^{3+}$ site, there are four short and two long Mn$-$O bonds with the average bond lengths of 1.93 and 2.17 {\AA}, respectively. Since the distortion is limited mainly to the neighboring O atoms of the resulted Mn$^{3+}$ or Mn$^{4+}$ ion, these polarons can be regarded as small polarons.\cite{Stoneham2007}

We find that the calculated formation energy of the polarons is as low as 0.32 eV ($\eta^{+}$) or 0.47 eV ($\eta^{-}$), in which $\eta^{+}$ is always energetically more favorable. It should be noted that these are additionally formed polarons, {\it i.e.}, they are considered as ``defects'' as compared to the perfect bulk material. The self-trapping energies of $\eta^{+}$ and $\eta^{-}$ are 0.82 and 1.02 eV, respectively, defined as the difference between the formation energy of the free hole or electron and that of the hole or electron polaron.\cite{Hoang2014} With these high self-trapping energies, the polarons are very stable in \ce{LiMn2O4}. This is not surprising, given the consideration that half of the Mn sites in the host compound can be regarded as being stable as hole polarons (Mn$^{4+}$) in a hypothetical all-Mn$^{3+}$ \ce{LiMn2O4} and the other half can be regarded as electron polarons (Mn$^{3+}$) in a hypothetical all-Mn$^{4+}$ \ce{LiMn2O4}.

{\bf Vacancies and interstitials.} The formation of $V_{\rm Li}^-$ involves removing a Li$^{+}$ ion, which causes negligible disturbance in the lattice. $V_{\rm Li}^0$ is, on the other hand, created by removing a Li atom, which is in fact a Li$^{+}$ ion and an electron from a neighboring Mn atom. This results in a void at the site of the removed Li$^{+}$ and a Mn$^{4+}$ at a neighboring Mn site (originally a Mn$^{3+}$ ion). $V_{\rm Li}^0$ is thus not the neutral charge state of a lithium vacancy but a complex of $V_{\rm Li}^-$ and $\eta^{+}$. This defect complex has a binding energy of 0.60 eV with respect to $V_{\rm Li}^-$ and $\eta^{+}$. Defects such as $V_{\rm Li}^-$, as well as $\eta^{+}$ and $\eta^{-}$, are regarded as {\it elementary} defects; the structure and energetics of other defects, {\it e.g.}, $V_{\rm Li}^0$, can be interpreted in terms of these basic building blocks. We find that the formation energy of $V_{\rm Li}^0$ is always lower than that of $V_{\rm Li}^-$. For lithium interstitials, Li$_{i}^{+}$ is created by adding a Li$^{+}$ ion. The defect is energetically most favorable when combining with another Li$^{+}$ ion from an 8a site to form a Li$-$Li dumbbell centered at the 8a site. This is in contrast to what has been commonly assumed that lithium interstitials are most stable at the 16c sites. The energy in the dumbbell configuration is lower than that at the 16c site by at least 0.15 eV. Finally, Li$_{i}^{0}$, created by adding a Li atom, is a complex of Li$_{i}^{+}$ and $\eta^{-}$ with a binding energy of 0.52 eV.

Among the manganese vacancies, $V_{\rm Mn}^{3-}$, {\it i.e.}, the removal of a Mn$^{3+}$ ion, is the elementary defect. Other defects such as $V_{\rm Mn}^{2-}$, $V_{\rm Mn}^{-}$, or $V_{\rm Mn}^{0}$ are complexes of $V_{\rm Mn}^{3-}$ and, respectively, one, two, or three $\eta^{+}$. $V_{\rm Mn}^{0}$ is found to be the lowest-energy manganese vacancy configuration, suggesting that it is more favorable to form a manganese vacancy when it is surrounded by Mn$^{4+}$ ions. The removal of manganese from the Mn$^{3+}$ site costs less energy than from the Mn$^{4+}$ site; the formation energy difference is about 0.20 eV or higher. Regarding the oxygen vacancies, the removal of an O$^{2-}$ ion does not lead to a void formed by the removed ion, often denoted as $V_{\rm O}^{2+}$, but a complex of $V_{\rm O}^{2+}$ and a hole-electron polaron pair ($\eta^{+}$$-$$\eta^{-}$), hereafter denoted as $V_{\rm O*}^{2+}$. Clearly, $V_{\rm O}^{2+}$ is not stable as a single point defect, and its formation is associated with some Mn$^{3+}$/Mn$^{4+}$ disorder. Other oxygen vacancies such as $V_{\rm O}^{+}$ or $V_{\rm O}^{0}$ are complexes of $V_{\rm O}^{2+}$ and one or two $\eta^{-}$. We find that the oxygen vacancies have much higher formation energies than other intrinsic defects; their lowest value is 1.49 eV for $V_{\rm O}^{0}$ at point C in the chemical-potential diagram. We also investigated manganese and oxygen interstitials and found that they all have very high formation energies (about 3 eV or higher), suggesting that these interstitials are not likely to form in the material.

\begin{figure}
\centering
\includegraphics[width=3.2in]{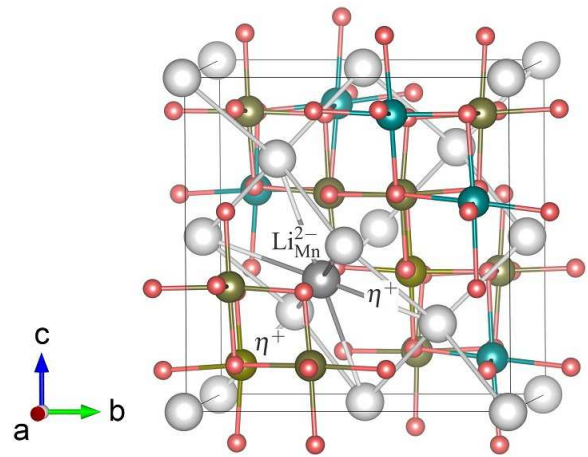}
\caption{Structure of Li$_{\rm Mn}^{0}$, the intrinsic ionic defect with the lowest formation energy in \ce{LiMn2O4}. This neutral defect is a complex of one negatively charged antisite Li$_{\rm Mn}^{2-}$ (large dark gray sphere) and two small hole polarons $\eta^{+}$ (medium light yellow spheres).}
\label{fig;limn10n}
\end{figure}

{\bf Antisite defects.} Lithium antisites Li$_{\rm Mn}$ are created by replacing Mn at a Mn site with Li. Li$_{\rm Mn}^{2-}$, {\it i.e.}, Li$^{+}$ substituting Mn$^{3+}$, is an elementary defect. Other antisites such as Li$_{\rm Mn}^{-}$ or Li$_{\rm Mn}^{0}$ are complexes of Li$_{\rm Mn}^{2-}$ and, respectively, one or two $\eta^{+}$. Among all possible ionic defects in \ce{LiMn2O4}, Li$_{\rm Mn}^{0}$ has the lowest formation energy, as low as 0.11 eV. Figure~\ref{fig;limn10n} shows the structure of Li$_{\rm Mn}^{0}$ where Li$_{\rm Mn}^{2-}$ is clearly seen surrounded by six Mn$^{4+}$ ions ({\it i.e.}, $\eta^{+}$) and six Li$^{+}$ ions. It should be noted that the two $\eta^{+}$ in the Li$_{\rm Mn}^{0}$ complex are created together with the Li$_{\rm Mn}^{2-}$, in addition to those Mn$^{4+}$ ions already present in bulk \ce{LiMn2O4}. We find that the energy cost for ion substitution at the Mn$^{3+}$ site is lower than at the Mn$^{4+}$ site; the energy difference is 0.26 eV or higher. Manganese antisites Mn$_{\rm Li}$ are created in a similar way by replacing Li at a Li (tetrahedral) site with Mn. Mn$_{\rm Li}^{+}$ is an elementary defect, in which the Mn ion is stable as high-spin Mn$^{2+}$ with a calculated magnetic moment of 4.45 $\mu_{\rm B}$. Other manganese antisites such as Mn$_{\rm Li}^{0}$ or Mn$_{\rm Li}^{2+}$ are complexes of Mn$_{\rm Li}^{+}$ and $\eta^{-}$ or $\eta^{+}$. For comparison, in layered \ce{LiMO2} (M = Ni, Co) the transition metal is also found to be stable as high-spin M$^{2+}$ at the Li (octahedral) site.\cite{Hoang2014}

{\bf Defect complexes.} In addition to the above defects, we explicitly investigated hole-electron polaron pairs ($\eta^{+}$$-$$\eta^{-}$), antisite defect pairs (Mn$_{\rm{Li}}$$-$Li$_{\rm{Mn}}$), and lithium Frenkel pair (Li$_{i}^+$$-$$V_{\rm{Li}}^-$). The hole-electron polaron pair is formed by switching the positions of a Mn$^{3+}$ and its neighboring Mn$^{4+}$ ion. After structural relaxations, the pair distance is 2.93 {\AA}. This defect pair has a formation energy of 0.37 eV and a binding of 0.54 eV. The antisite pair is created by switching the positions of a Li atom and its neighboring Mn atom. This ultimately results in a Mn$_{\rm{Li}}^{+}$$-$Li$_{\rm{Mn}}^{2-}$$-$$\eta^{+}$ complex, in which the distance from the manganese antisite to the lithium antisite is 2.92 {\AA} and that from the lithium antisite to the hole polaron is 3.47 {\AA}. This complex has a formation energy of 0.63 eV and a binding of 1.76 eV. Finally, the lithium Frenkel pair is created by moving a Li$^{+}$ ion away from an 8a site to form a Li$-$Li dumbbell with another Li$^{+}$ ion. This results in a $V_{\rm{Li}}^-$ at the 8a site and a Li$_{i}^+$. After relaxations, the distance between the vacancy and the center of the dumbbell is about 4.0 {\AA} (The pair is unstable toward recombination at shorter distances). This pair has a formation energy of 1.85 eV and a binding energy of 0.30 eV. It should be noted that the formation energies of these three defect complexes are independent of the chemical potentials.

\subsection{Defect migration}

\begin{figure}
\centering
\includegraphics[width=3.2in]{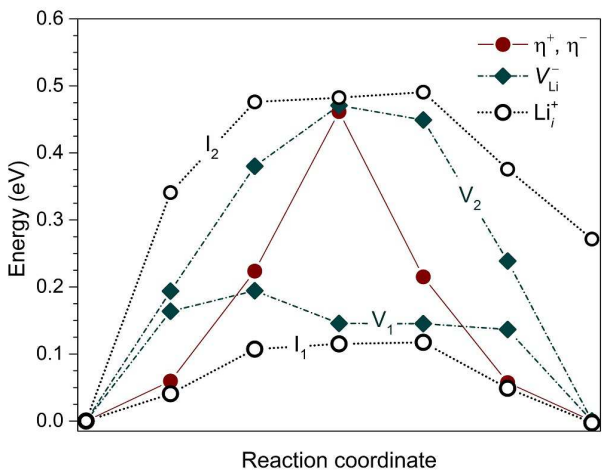}
\caption{Migration barriers of the small hole ($\eta^{+}$) and electron ($\eta^{-}$) polarons and lithium vacancies ($V_{\rm Li}^{-}$) and interstitials (Li$_{i}^{+}$) in long-range charge-ordered \ce{LiMn2O4}.}
\label{fig;migration}
\end{figure}

Figure~\ref{fig;migration} shows the migration barriers ($E_{m}$) for the hole and electron polarons and lithium vacancies and interstitials in \ce{LiMn2O4}. All the migration barrier calculations were carried out with the $\Gamma$ point only. The migration of a polaron between two positions \textbf{q$_{1}$} and \textbf{q$_{2}$} can be described by the transfer of the lattice distortion over a one-dimensional Born-Oppenheimer surface.\cite{Maxisch:2006p103} We estimate the energy barrier by computing the energies of a set of cell configurations linearly interpolated between \textbf{q$_{1}$} and \textbf{q$_{2}$} and identify the energy maximum. The migration barrier of $\eta^{+}$ and $\eta^{-}$ is found to be 0.46 eV. For comparison, Ouyang {\it et al.}\cite{Ouyang2009PLA} reported a migration barrier of 0.35 eV for polarons in \ce{LiMn2O4}, obtained in GGA$+U$ calculations with $U=4.5$ eV.

For lithium vacancies $V_{\rm Li}^{-}$, we find two distinct migration paths with barriers of 0.19 eV (path V$_{1}$) and 0.47 eV (path V$_{2}$), calculated using the NEB method.\cite{ci-neb} Both paths involve moving a Li$^{+}$ ion from an 8a site to the vacancy (an empty 8a site) through an interstitial 16c site. Here, moving a Li$^{+}$ ion in one direction is equivalent to $V_{\rm Li}^{-}$ migrating in the opposite direction. The migration bottleneck is a Mn ring at the 16c site, consisting of six Mn ions in the plane perpendicular to the migration path. In path V$_{1}$, the Mn ring has four Mn$^{4+}$ ions and two Mn$^{3+}$ ions, whereas in path V$_{2}$ it has two Mn$^{4+}$ ions and four Mn$^{3+}$ ions. We have also considered situations in which Li$^{+}$ ions migrate through a Mn ring that consists of three Mn$^{4+}$ ions and three Mn$^{3+}$ ions and find migration barriers of 0.47$-$0.57 eV. An example of such situations is when the $\eta^{+}$ component of the $V_{\rm Li}^{0}$ complex is kept fixed while the $V_{\rm Li}^{-}$ component of the complex is migrating. 

Lithium interstitials Li$_{i}^{+}$ migrate through an interstitialcy mechanism involving concerted motion of three lithium ions: two ions of the Li$-$Li dumbbell and one ion that is next to the dumbbell. We find barriers of 0.12 eV for the migration path (hereafter called path I$_{1}$) that goes through Mn rings all consisting of four Mn$^{4+}$ ions and two Mn$^{3+}$ ions, and 0.49 eV for the path (path I$_{2}$) that goes through at least one Mn ring that consists of two Mn$^{4+}$ ions and four Mn$^{3+}$ ions.

For comparison, Xu and Meng\cite{Xu2010} obtained from GGA$+U$ calculations with $U$ = 4.84 eV lithium migration barriers of $\sim$0.2$-$0.4 eV associated with Mn$^{4+}$-rich rings, $\sim$0.6 eV with Mn$^{3+}$-rich rings, and $\sim$0.8 eV with Mn rings that have equal numbers of Mn$^{4+}$ and Mn$^{3+}$ ions. The migrating species in their calculations could be $V_{\rm Li}^{0}$, instead of $V_{\rm Li}^{-}$ like in our calculations. However, it is not clear from their work how the two components of the $V_{\rm Li}^{0}$ complex migrate relative to each other. We note that, in our calculations, a lower bound on the migration barrier of a defect complex, {\it e.g.}, $V_{\rm Li}^{0}$ or Li$_{i}^{0}$, can be estimated by taking the higher of the migration barriers of its constituents.\cite{wilson-short} 

We find that the tetragonal distortion has minor effects on the migration barriers. For example, our calculations using cubic supercells that have the same volume and Mn$^{3+}$/Mn$^{4+}$ arrangement as the tetragonal supercells give $V_{\rm Li}^{-}$ barriers of 0.20 eV and 0.45 eV for paths V$_{1}$ and V$_{2}$, respectively, which are almost identical to the values reported earlier. However, in the presence of Mn$^{3+}$/Mn$^{4+}$ disorder and other lattice defects, {\it e.g.}, in lithium over-stoichiometric Li[Mn$_{2-\alpha}$Li$_{\alpha}$]O$_{4}$, the lithium ions are not likely to encounter Mn rings that are all Mn$^{4+}$-rich for the whole diffusion length, and the overall migration barrier will therefore be determined by the higher-barrier segment. As a result, the hole and electron polarons and lithium vacancies and interstitials, except the lithium at the octahedral 16d site, all have an estimated migration barrier of about 0.5 eV.

Finally, we investigated the migration of the Li$^{+}$ ion that is associated with the lithium antisite Li$_{\rm Mn}^{0}$, {\it cf.}~Fig.~\ref{fig;limn10n}. The energy barrier for Li$^{+}$ migration from the octahedral 16d site to one of the six neighboring Li tetrahedral 8a sites through a vacancy mechanism is found to be 0.6$-$2.0 eV. With this higher migration barrier, the ion is trapped at the 16d site and expected not be deintercalated during charging. 

\section{Discussion}

We list in Table~\ref{tab:formenergy} the calculated formation energies of relevant point defects in \ce{LiMn2O4} for three different sets of the atomic chemical potentials, corresponding to different sets of the experimental conditions. The chemical potential of oxygen, $\mu_{\rm O}$, is $-$0.15 eV, $-$0.42 eV, and $-$0.69 eV, respectively, at points A, B, and C in the chemical-potential diagram, {\it cf.}~Fig.~\ref{fig;chempot}. $\mu_{\rm O}$ can be controlled by controlling temperature and pressure and/or oxygen reducing agents. Lower $\mu_{\rm O}$ values are usually associated with higher temperatures and/or lower oxygen partial pressures and/or the presence of oxygen reducing agents. For each set of the atomic chemical potentials, the formation energy values are obtained at the respective Fermi-level position $\mu_{e}^{\rm int}$. We find that $\mu_{e}^{\rm int}$ is at 0.98$-$1.11 eV, which is always away from both the VBM and CBM. Most of the defects have a calculated formation energy of 1.0 eV or lower, at least under certain conditions. They can therefore occur in the material with high concentrations, {\it e.g.}, during synthesis. These defects, except the mobile ones such as the polarons and lithium vacancies, are expected to get trapped when the material is cooled to room temperature. We also find that the formation energy of the polaron pair $\eta^{+}$$-$$\eta^{-}$ is low, only 0.37 eV, indicating that \ce{LiMn2O4} is prone to Mn$^{3+}$/Mn$^{4+}$ disorder. Mn$_{\rm{Li}}$$-$Li$_{\rm{Mn}}$ also has a low formation energy, which suggests the presence of cation mixing (see further discussions in Sections 4.3 and 4.4).

\subsection{Electronic and ionic conduction}

\begin{table}
\caption{\ Formation energies ($E^{f}$) and binding energies ($E_{b}$) of intrinsic defects in \ce{LiMn2O4}. The formation energies are obtained at points A, B, and C in the chemical-potential diagram}\label{tab:formenergy}
\small
\begin{tabular*}{0.47\textwidth}{@{\extracolsep{\fill}}lllllr}
\hline
Defect&\multicolumn{3}{c}{$E^{f}$ (eV)}&Constituents&$E_{b}$ (eV) \\
\hline
&A&B&C&& \\
\hline
$\eta^{+}$&0.32&0.32&0.45&\\ 
$\eta^{-}$ &0.59&0.59&0.47&\\ 
$\eta^{+}$$-$$\eta^{-}$&0.37&0.37&0.37&$\eta^{+} + \eta^{-}$&0.54\\
$V_{\rm{Li}}^-$&0.92&1.28&1.33&\\ 
$V_{\rm{Li}}^0$&0.64&1.00&1.18&$V_{\rm{Li}}^- + \eta^{+}$&0.60\\
Li$_{i}^+$&1.23&0.87&0.82&\\
Li$_{i}^0$&1.30&0.94&0.76&Li$_{i}^+ + \eta^{-}$&0.52\\
Li$_{i}^+$$-$$V_{\rm{Li}}^-$&1.85&1.85&1.85&Li$_{i}^+ + V_{\rm{Li}}^-$&0.30\\
Li$_{\rm{Mn}}^{2-}$&1.30&1.30&1.31&\\ 
Li$_{\rm{Mn}}^-$&0.32&0.32&0.46&Li$_{\rm{Mn}}^{2-} + \eta^{+}$&1.30\\
Li$_{\rm{Mn}}^0$&0.11&0.11&0.38&Li$_{\rm{Mn}}^{2-} + 2\eta^{+}$&1.82\\
Mn$_{\rm{Li}}^+$&0.77&0.77&0.62&\\
Mn$_{\rm{Li}}^0$&0.92&0.92&0.65&Mn$_{\rm{Li}}^+ + \eta^{-}$&0.44\\
Mn$_{\rm{Li}}^{2+}$&1.02&1.02&1.00&Mn$_{\rm{Li}}^+ + \eta^{+}$&0.07\\ 
Mn$_{\rm{Li}}$$-$Li$_{\rm{Mn}}$&0.63&0.63&0.63&Mn$_{\rm{Li}}^{+} + \rm{Li}_{\rm{Mn}}^{2-} + \eta^{+}$&1.76\\
$V_{\rm{Mn}}^{3-}$&3.11&3.47&3.51&\\ 
$V_{\rm{Mn}}^{2-}$&1.80&2.16&2.34&$V_{\rm{Mn}}^{3-} + \eta^{+}$&1.63\\
$V_{\rm{Mn}}^{-}$&1.03&1.39&1.71&$V_{\rm{Mn}}^{3-} + 2\eta^{+}$&2.72\\
$V_{\rm{Mn}}^{0}$&0.80&1.16&1.61&$V_{\rm{Mn}}^{3-} + 3\eta^{+}$&3.27\\
$V_{\rm{O*}}^{2+}$&2.42&2.15&2.15&$V_{\rm{O}}^{2+} + \eta^{+} + \eta^{-}$\\ 
$V_{\rm{O}}^{+}$&2.15&1.88&1.75&$V_{\rm{O}}^{2+} + \eta^{-}$&\\
$V_{\rm{O}}^{0}$&2.03&1.76&1.49&$V_{\rm{O}}^{2+} + 2\eta^{-}$&\\
\hline \\
\end{tabular*}
\end{table}

Strictly speaking, each ionic defect in \ce{LiMn2O4} has only one stable charge state, which is also called the elementary defect; oxygen vacancies do not even have any configuration that is stable as a single point defect, as mentioned earlier. Removing (adding) electrons from (to) these elementary defects always results in defect complexes consisting of the elementary defects and small hole (electron) polarons. Besides, several positively and negatively charged defects have positive formation energies only near midgap, {\it cf.}~Fig.~\ref{fig;formenergy}, making them perfect charge-compensators. Any attempt to deliberately shift the Fermi level of the system from $\mu_{e}^{\rm int}$ to the VBM or CBM will result in the charged defects having negative formation energies, {\it i.e.}, the intrinsic defects will form spontaneously and counteract the effects of shifting.\cite{Hoang2011,Hoang2014} Clearly, intrinsic point defects in \ce{LiMn2O4} cannot act as sources of band-like electrons and holes, and the material cannot be made n-type or p-type. The electronic conduction therefore proceeds via hopping of small hole and electron polarons. Regarding the ionic conduction, lithium ions are the current-carrying species, which migrate via vacancy and/or interstitialcy mechanisms. 

The activation energies for electronic and ionic conduction can be estimated from the formation energies and migration barriers of the current-carrying defects, $E_{a} = E^{f} + E_{m}$. As discussed earlier, except for paths V$_{1}$ and I$_{1}$ which are unlikely to be realized in Mn$^{3+}$/Mn$^{4+}$-disordered \ce{LiMn2O4}, the barriers for the polarons and lithium ions are basically similar, $\sim$0.5 eV. Therefore, the relative contribution of a defect or migration mechanism to the total conductivity is determined exclusively by the defect's concentration. If the defect is predominantly {\it athermal}, as it is the case for hole and electron polarons in \ce{LiMn2O4} and lithium vacancies in Li-deficient or partially delithiated Li$_{1-x}$Mn$_{2}$O$_{4}$, the activation energy includes only the migration part, {\it i.e.}, $E_{a} = E_{m}$.\cite{Hoang2014} The electronic activation energy is thus $\sim$0.5 eV, {\it i.e.}, the barrier for polarons. In partially delithiated Li$_{1-x}$Mn$_{2}$O$_{4}$, the ionic activation energy is also $\sim$0.5 eV, {\it i.e.}, the barrier for lithium vacancies. In stoichiometric \ce{LiMn2O4}, however, lithium vacancies and/or interstitials have to be thermally activated; the ionic activation energy includes both the migration ($\sim$0.5 eV) and formation ({\it cf.}~Table~\ref{tab:formenergy}) parts, which is estimated to be as low as $\sim$1.1 eV (1.3 eV) for the lithium vacancy (interstitialcy) mechanism. Clearly, the total conductivity is dominated by the electronic contribution.

Experimentally, Fang and Chung\cite{Fang2008} reported activation energies of 0.43 eV and 0.38 eV for electronic conduction in \ce{LiMn2O4} below and above room temperature, respectively. Other authors reported values of 0.40$-$0.44 eV.\cite{Massarotti1999,Iguchi2002} Regarding lithium diffusion, Verhoeven {\it et al.}\cite{Verhoeven2001} obtained an activation energy of 0.5$\pm$0.1 eV in the temperature range 345$-$400 K from $^{7}$Li NMR experiments on Li[Mn$_{1.96}$Li$_{0.04}$]O$_{4}$. Takai {\it et al.}\cite{Takai2014} reported activation energies of 0.52 eV and 1.11 eV for lithium diffusion in \ce{LiMn2O4} below and above 600$^{\circ}$C, extracted from tracer diffusion coefficients measured by neutron radiography. The lower-temperature values are very close to our estimated migration barrier. The value 1.11 eV at high temperatures could indicate that the system is in the intrinsic region where the activation energy includes both the formation and migration parts. 

\subsection{Delithiation and lithiation} 

The structure of the lithium vacancy $V_{\rm Li}^0$ in a battery electrode material often provides direct information on the delithiation mechanism. In \ce{LiMn2O4}, $V_{\rm Li}^0$ indicates that for each Li atom removed from \ce{LiMn2O4} electrodes during delithiation, the material is left with one negatively charged lithium vacancy $V_{\rm Li}^-$ and one hole polaron $\eta^{+}$; {\it i.e.}, the extraction of Li is associated with the oxidation of Mn$^{3+}$ to Mn$^{4+}$. The deintercalation voltage\cite{Aydinol1997} associated with the extraction of the first lithium, {\it i.e.}, the creation of $V_{\rm Li}^0$, is 4.29 V. The partially delithiated composition can be written as Li$_{1-x}$Mn$_{2}$O$_{4}$ (Here we ignore the pre-existing intrinsic defects which will be discussed later). The lithium interstitial Li$_{i}^{0}$, on the other hand, provides information on the lithiation mechanism. For each Li atom inserted into \ce{LiMn2O4} electrodes during lithiation, the material receives one positively lithium interstitial Li$_{i}^+$ and one electron polaron $\eta^{-}$; {\it i.e.}, the Li insertion is associated with the reduction of Mn$^{4+}$ to Mn$^{3+}$. The partially lithiated composition can be written as Li$_{1+x}$Mn$_{2}$O$_{4}$ (Not to be confused with lithium over-stoichiometric Li$_{1+\alpha}$Mn$_{2-\alpha}$O$_{4}$ where the Li excess replaces Mn at the 16d sites). Since there are no band-like carriers, $\eta^{+}$ and $\eta^{-}$ are the electronic charge carriers in the delithiation and lithiation processes. Also, it is important to note that polarons and lithium vacancies (interstitials) created from delithiation (lithiation) are not thermal defects. 

\subsection{Lithium over-stoichiometry} 

Among all possible ionic defects in \ce{LiMn2O4}, the lithium antisite Li$_{\rm Mn}^{0}$ is dominant. The defect has a formation energy of 0.11$-$0.38 eV, depending the specific set of the atomic chemical potentials. For comparison, the calculated formation energy of the lithium antisite in layered \ce{LiCoO2} and \ce{LiNiO2} is in the range of 0.92$-$2.73 eV (Li$_{\rm Co}^{0}$) or 0.68$-$1.11 eV (Li$_{\rm Ni}^{0}$).\cite{Hoang2014} The low energy of Li$_{\rm Mn}^{0}$ can partially be ascribed to the small difference in the ionic radii of Li$^{+}$ (0.76 {\AA}) and high-spin Mn$^{3+}$ (0.65 {\AA}); for reference, the Shannon ionic radii of six-fold coordinated, low-spin Co$^{3+}$ and Ni$^{3+}$ are 0.55 {\AA} and 0.56 {\AA}, respectively.\cite{Shannon1976} Given Li$_{\rm Mn}^{0}$ with that low formation energy, the synthesis of \ce{LiMn2O4} under equilibrium or near-equilibrium conditions is expected to result in a lithium over-stoichiometric compound with the composition Li$_{1+\alpha}$Mn$_{2-\alpha}$O$_{4}$ or, more explicitly, Li[Mn$_{2-\alpha}$Li$_{\alpha}$]O$_{4}$ or Li$^{+}$[Mn$_{1-3\alpha}^{3+}$Mn$_{1+2\alpha}^{4+}$Li$_{\alpha}^{+}$]O$_{4}^{2-}$. In this composition, each negatively charged lithium antisite Li$_{\rm{Mn}}^{2-}$ is charge-compensated by two hole polarons $\eta^{+}$, and the average Mn oxidation state is higher than +3.5; {\it i.e.}, Mn$^{4+}$ is slightly more favorable than Mn$^{3+}$. Since the Li$^{+}$ ion at the octahedral 16d site gets trapped due to its lower mobility, it is unlikely to be deintercalated during charging. Besides, Li[Mn$_{2-\alpha}$Li$_{\alpha}$]O$_{4}$ has only ($1-3\alpha$) Mn$^{3+}$ ions for the oxidation reactions. As a result, there will be residual lithium in the fully delithiated compound, both at the 16d and 8a sites, {\it i.e.}, Li$_{3\alpha}^{+}$[Mn$_{2-\alpha}^{4+}$Li$_{\alpha}^{+}$]O$_{4}^{2-}$. The theoretical capacity will therefore decrease from 148 mAh/g to $148(1-3\alpha)$ mAh/g.

Our results for Li$_{\rm Mn}^{0}$ thus explain why \ce{LiMn2O4} samples are often lithium over-stoichiometric. Martinez {\it et al.}\cite{Martinez2014} found about 90\% of the lithium ions at the tetrahedral 8a sites and 10\% at the octahedral 16d sites, confirming that the Li excess replaces Mn at the 16d sites. Xia and Yoshio\cite{Xia1996} seemed to indicate that stoichiometric \ce{LiMn2O4} electrodes are unstable toward the lithium over-stoichiometric ones. We note that, as a consequence of the lithium over-stoichiometry and the likely random distribution of Li$_{\rm Mn}^{0}$, the transformation from the Mn$^{3+}$/Mn$^{4+}$-disordered, cubic phase to a long-range ordered, tetragonal/orthorhombic phase may not be realized in practice, even at low temperatures. In fact, there have been reports of the absence of the long-range charge order associated with the Jahn-Teller effect in lithium over-stoichiometric Li$_{1+\alpha}$Mn$_{2-\alpha}$O$_{4}$.\cite{Lee2002,Kamazawa2011} Kamazawa {\it et al.},\cite{Kamazawa2011} for example, observed only short-range charged-order in Li$_{1.1}$Mn$_{1.9}$O$_{4}$. Regarding the electrochemical performance, Li$_{1+\alpha}$Mn$_{2-\alpha}$O$_{4}$ samples have been reported to show a significantly enhanced cycling stability compared to stoichiometric \ce{LiMn2O4}, although at the expense of capacity.\cite{Xia1996,Gummow1994} The enhancement has been attributed mainly to the suppression of the Jahn-Teller effect on deep discharge. However, the presence of the residual lithium in the delithiated compound, {\it i.e.}, Li$_{3\alpha}^{+}$[Mn$_{2-\alpha}^{4+}$Li$_{\alpha}^{+}$]O$_{4}^{2-}$, can also help improve the cycling stability, unlike in \ce{LiMn2O4} where the complete extraction of lithium results in unstable $\lambda$-MnO$_{2}$ electrodes.\cite{Gummow1994}
 
\subsection{Other possible defects}

The manganese antisite Mn$_{\rm Li}^{+}$ can occur in \ce{LiMn2O4}, given its formation energy of only 0.62$-$0.77 eV. For comparison, the formation energy of Co$_{\rm Li}^{+}$ in \ce{LiCoO2} is 0.55$-$2.08 eV and that of Ni$_{\rm Li}^{+}$ in \ce{LiNiO2} is 0.53$-$0.96 eV.\cite{Hoang2014} Mn$_{\rm Li}^{+}$ can be created together with Li$_{\rm Mn}^{2-}$ and $\eta^{+}$, {\it i.e.}, in form of Mn$_{\rm{Li}}$$-$Li$_{\rm{Mn}}$ with a formation energy of 0.63 eV, or with $\eta^{-}$, {\it i.e.}, in form of Mn$_{\rm Li}^{0}$ with a formation energy of 0.65$-$0.92 eV, {\it cf.}~Table~\ref{tab:formenergy}. Manganese antisites have also been observed in experiments. Bj{\"{o}}rk {\it et al.}\cite{Bjork2001} reported that 9\% of the lithium ions at the tetrahedral 8a sites were substituted by Mn$^{2+}$ ions in high-temperature synthesis. This is consistent with our results showing that Mn$_{\rm Li}^{+}$ has the lowest formation energy at point C which corresponds to a low $\mu_{\rm O}$ value. If created in form of Mn$_{\rm Li}^{0}$, the co-generation of $\eta^{-}$ will result in an increase in the amount of the Jahn-Teller distorted Mn$^{3+}$ ions. Besides, we speculate that manganese antisites may also act as nucleation sites for the formation of impurity phases during electrochemical cycling. 

Next, with a formation energy of 0.76$-$1.23 eV, lithium interstitials can also occur in the material, {\it e.g.}, when synthesized under conditions near point C in the chemical-potential diagram, {\it cf.}~Fig.~\ref{fig;chempot}. Li$_{i}^{+}$ can be created together with $\eta^{-}$, {\it i.e.}, in form of Li$_{i}^{0}$ which results in the composition Li$_{1+\alpha}$Mn$_{2}$O$_{4}$ (assuming no other defects). In this composition, the average Mn oxidation state is $<$3.5. This defect is unlikely to form through the lithium Frenkel pair mechanism, {\it i.e.}, in combination with $V_{\rm{Li}}^-$, because the Li$_{i}^+$$-$$V_{\rm{Li}}^-$ pair is either unstable or high in energy. Experimentally, Berg {\it et al.}\cite{Berg1999} reported that, in their Li$_{1+\alpha}$Mn$_{2-\alpha}$O$_{4}$ ($\alpha = 0.14$) samples, lithium ions occupy both the 8a sites with 100\% occupancy and the 16c sites with 7.0\% occupancy, and manganese ions occupy the 16d sites with 93.0\% occupancy. In light of our results for the lithium interstitials, it would be interesting to re-examine the samples and see if some of the lithium is really stable at the 16c sites. The results of Berg {\it et al.}~may also suggest the presence of Li$_{i}$ and $V_{\rm{Mn}}$ in form of a neutral Li$_{i}$$-$$V_{\rm{Mn}}$ complex. However, we find that this complex has a high formation energy (1.93$-$2.20 eV), indicating that it is not likely to occur under equilibrium or near-equilibrium synthesis conditions.

Manganese vacancies can form when synthesized at lower temperatures, {\it e.g.}, at point A in Fig.~\ref{fig;chempot} where the formation energy of $V_{\rm Mn}^{0}$ is just 0.80 eV, {\it cf.}~Table~\ref{tab:formenergy}. Gummow {\it et al.}\cite{Gummow1994} reported the presence of vacancies on both the 8a and 16d sites in Li$_{1-\alpha}$Mn$_{2-2\alpha}$O$_{4}$ ($0<\alpha\leq$0.11) synthesized at temperatures between 400 and 600$^\circ$C, although they also acknowledged that samples with a precise, predetermined composition were difficult to prepare. The cycling stability of this compound was found to be inferior to that of Li$_{1+\alpha}$Mn$_{2-\alpha}$O$_{4}$.\cite{Gummow1994} Finally, with a much higher formation energy (1.49$-$2.03 eV), oxygen vacancies are expected not to occur inside the material. This is consistent with experiments where no oxygen vacancies have been found.\cite{Paulsen1999,Bjork2001} We note that oxygen vacancies may still occur at the surface or interface where the lattice environment is less constrained than in the bulk. The formation of manganese vacancies, as well as lithium interstitials and manganese antisites, is also expected to be energetically more favorable at the surface or interface.

Apparently, manganese antisites and vacancies and lithium interstitials can lead to inferior cycling stability and hence should be avoided. One can tune the synthesis conditions to reduce these defects in \ce{LiMn2O4} samples. From our results summarized in Table~\ref{tab:formenergy}, the best compromise could be to synthesize the material under the conditions near point B in the chemical-potential diagram where there is a three-phase equilibrium between \ce{LiMn2O4}, \ce{Li2MnO3} and \ce{Li5Mn7O16}, {\it cf.}~Fig.~\ref{fig;chempot}. The formation energy of the manganese antisites under these conditions is, however, still quite low (0.77 eV). Further reduction of the defects may thus require partially ion substitution that can significantly change the chemical environment and hence the defect landscape.

\section{Conclusions}

We have carried out a DFT study of the bulk properties and defect thermodynamics and transport in spinel-type \ce{LiMn2O4}, using the HSE06 screened hybrid density functional. We find that the tetragonal distortion of cubic \ce{LiMn2O4} during structural optimizations at 0 K is associated with charge ordering. The compound is found to be thermodynamically stable and its stability region in the Li$-$Mn$-$O phase diagram is defined by the \ce{Mn2O3}, \ce{Li2MnO3}, and \ce{Li5Mn7O16} phases.

Intrinsic electronic and ionic defects in \ce{LiMn2O4} can form with high concentrations. Several charged defects have positive formation energies only in a region near midgap, making them perfect charge-compensators. The defects cannot act as sources of band-like carriers and the material cannot be doped n- or p-type. The electronic conduction proceeds via hopping of the small hole and electrons polarons and the ionic conduction occurs via lithium vacancy and/or interstitialcy migration mechanisms. The total bulk conductivity is found to be predominantly from the electronic contribution. An analysis of the structure of lithium vacancies and interstitials shows that lithium extraction (insertion) is associated with the oxidation (reduction) reaction at the Mn site.

Among the intrinsic ionic defects, lithium antisites are the dominant defect with a very low calculated formation energy. This low energy is ascribed to the small ionic radius difference between Li$^{+}$ and high-spin Mn$^{3+}$. The formation energy of the hole-electron polaron pair is also very low. Our results thus indicate that \ce{LiMn2O4} is prone to lithium over-stoichiometry and Mn$^{3+}$/Mn$^{4+}$ disorder. In the lithium over-stoichiometric compound, there is residual lithium that is not deintercalated during charging and can help improve the cycling stability. Other defects such as manganese antisites and vacancies and lithium interstitials can also occur, under certain experimental conditions, but with lower concentrations than the lithium antisites. An elimination of the manganese antisites may require significant changes to the chemical environment, {\it e.g.}, through ion substitution.  

\begin{acknowledgments}

This work was supported by the U.S.~Department of Energy (Grant No.~DE-SC0001717) and the Center for Computationally Assisted Science and Technology (CCAST) at North Dakota State University.

\end{acknowledgments}


\footnotesize{
\providecommand*\mcitethebibliography{\thebibliography}
\csname @ifundefined\endcsname{endmcitethebibliography}
  {\let\endmcitethebibliography\endthebibliography}{}

}
\end{document}